\documentclass{article}

\usepackage{amsmath,amsthm,amsfonts,braket,graphicx}

\newcommand{\ketbra}[1]{\ket{#1}\bra{#1}}

\title{An Improved Asymptotic Key Rate Bound for a Mediated Semi-Quantum Key Distribution Protocol}
\author{Walter O. Krawec\\\small{Iona College}\\\small{New Rochelle, NY 10801 USA}\\\small{\texttt{walter.krawec@gmail.com}}}

\begin{document}
\maketitle

\begin{abstract}
Semi-quantum key distribution (SQKD) protocols allow for the establishment of a secret key between two users Alice and Bob, when one of the two users (typically Bob) is limited or ``classical'' in nature.  Recently it was shown that protocols exists when both parties are limited/classical in nature if they utilize the services of a quantum server.  These protocols are called mediated SQKD protocols.  This server, however, is untrusted and, in fact, adversarial.  In this paper, we reconsider a mediated SQKD protocol and derive a new proof of unconditional security for it.  In particular, we derive a new lower bound on its key rate in the asymptotic scenario.  Furthermore, we show this new lower bound is an improvement over prior work, thus showing that the protocol in question can tolerate higher rates of error than previously thought.
\end{abstract}

\section{Introduction}
Quantum key distribution (QKD) protocols are designed to allow two users, Alice ($A$) and Bob ($B$), to establish a shared secret key, secure against even an all-powerful adversary Eve ($E$).  Since the creation, in 1984, of the BB84 protocol \cite{QKD-BB84}, there have been several protocols developed which achieve this end.  For a general survey, the reader is referred to \cite{QKD-survey}.

Semi-quantum key distribution (SQKD) protocols, first introduced in 2007 by Boyer et al., \cite{SQKD-first}, have the same goal: the establishment of a secret key, secure against an all-powerful adversary.  However now, instead of allowing both $A$ and $B$ to manipulate quantum resources (e.g., prepare and measure qubits in a variety of bases) as is permissible in a typical QKD protocol, only $A$ is allowed such liberties while $B$ is limited to performing certain ``classical'' or ``semi-quantum'' operations (what operations $B$ is limited to are discussed shortly).  In this scenario, $A$ is called the \emph{quantum user} while $B$ is called the \emph{classical user} (in a fully quantum protocol, such as BB84\cite{QKD-BB84}, both $A$ and $B$ are fully quantum).  Such protocols are theoretically interesting as they attempt to answer the question ``how quantum does a protocol need to be in order to gain an advantage over a classical one?'' \cite{SQKD-first,SQKD-second}

These SQKD protocols, by necessity, rely on a two-way quantum communication channel - one which permits a qubit to travel from the quantum user $A$, to the classical user $B$, then back to $A$.  The limited classical $B$, upon receiving a qubit, is able to do one of two things: he may \emph{measure and resend} the qubit or \emph{reflect} the qubit.  Measuring and resending involves taking the qubit received from $A$ and subjecting it to a $Z$ basis measurement (the $Z$ basis being $\{\ket{0}, \ket{1}\}$).  His result $\ket{r}$, for $r \in \{0,1\}$ is then resent to $A$.  Reflecting the qubit involves $B$ allowing the qubit to simply pass through his lab undistributed in which case he learns nothing about its state.  Thus, the classical user is only able to work directly with the computational $Z$ basis - he cannot, for example, perform a measurement in the Hadamard $X$ basis (denoted $\{\ket{\pm} = 1/\sqrt{2}(\ket{0}\pm\ket{1})\}$).

This two-way quantum channel, of course, greatly complicates the security analysis of these protocols as the attacker $E$ is allowed two opportunities to interact with the traveling qubit.  While several SQKD protocols have been proposed \cite{SQKD-first,SQKD-second,SQKD-3,SQKD-4,SQKD-Single-Security,SQKD-cl-A,SQKD-random-prepare}, up until recently, the bulk of security proofs for such semi-quantum protocols have focused on the notion of \emph{robustness}.  This property, introduced in \cite{SQKD-first,SQKD-second}, requires that any attack which causes $E$ to potentially gain information, by necessity causes a disturbance which $A$ and $B$, with non-zero probability, may detect.  Note that robustness says nothing about the amount of information gained (which may be high) nor the probability of detection (which may be low).

Recently, some work has been accomplished moving beyond robustness.  In particular \cite{SQKD-information} derived expressions relating the disturbance of $E$'s attack to her information gain for the SQKD protocol of \cite{SQKD-first}, assuming she is limited to performing \emph{individual attacks} (those attacks where $E$ is limited to performing the same attack operation each iteration of the protocol and is forced to measure her quantum memory before $A$ and $B$ utilize their key for any purpose - these are weaker attacks than \emph{collective attacks} - where $E$ performs the same operation each iteration but can postpone her measurement until any future time of her choice - and \emph{general attacks} - those attacks where $E$ is allowed to do anything within the laws of physics; the reader is referred to \cite{QKD-survey} for more information on these possible attack models).  Also, \cite{SQKD-cl-A} described a new SQKD protocol and computed a similar relation between the disturbance and information gain of $E$'s attack, though again assuming individual attacks.

Our work recently has been in the proof of unconditional security (making no assumptions on the type of attack $E$ uses) of several SQKD protocols.  In particular, in \cite{SQKD-Krawec-SecurityProof}, we proved the security of Boyer et al.,'s original SQKD protocol \cite{SQKD-first} showing that $A$ and $B$ are able to distill a secure secret key so long as the noise in the channel is less than $5.34\%$.  We have also, in \cite{SQKD-Single-Security}, derived a series of security results for single state SQKD protocols (single state protocols were introduced in \cite{SQKD-lessthan4}, though without proofs of unconditional security).  While that particular work stopped short of unconditional security proofs, the security results we prove there can be applied toward that goal \cite{SQKD-Krawec-dissertation}.

Finally, and also the topic of this current paper, in \cite{SQKD-MultiUser}, we designed a new \emph{mediated SQKD} protocol.  Such a protocol allows two limited classical users $A$ and $B$ to establish a secure secret key with the help of a quantum server (denoted $C$ for ``Center'').  With such a system, one could envision a QKD infrastructure consisting of several limited ``classical'' users, utilizing the services of this quantum server in order to distill secret keys (each key known only to a pair of classical users, not the server).  If the quantum server $C$ were honest, such a goal would be trivial.  Instead, we assumed the server is untrusted - indeed, we can assume the server is the all-powerful quantum adversary.

In our original work in \cite{SQKD-MultiUser}, we proved our mediated protocol's unconditional security by computing a lower bound on its key rate (to be defined shortly, roughly speaking, the key rate is the ratio of the number of secret key bits to the number of qubits sent) in the asymptotic scenario (as the number of qubits sent approaches infinity).  We showed that, if the server is ``semi-honest,'' that is the server follows our protocol correctly but afterwards tries to learn something about the key, then $A$ and $B$ may distill a secret key so long as the noise in the channel is less than $19.9\%$.  If the server is fully adversarial (that is, the server does not follow the protocol and instead performs any operations he likes within the laws of quantum physics), then $A$ and $B$ may distill a secret key so long as the noise in the channel is less than $10.65\%$ (a number that is close to BB84's $11\%$ \cite{QKD-renner-keyrate}).

In this paper, we reconsider our security proof and improve on these tolerated noise levels by computing a new bound on the protocol's key rate using an alternative method of proof.  In our original proof, we utilized a bound on the Jensen-Shannon Divergence from \cite{QC-info-trace-bound} to find bounds on the quantum mutual information between $A$ and $C$'s system.  Here, we will use a technique adapted from \cite{QKD-keyrate-general} (which we also successfully applied in our proof of security for a different SQKD protocol in \cite{SQKD-Krawec-SecurityProof}) which allows us to bound the conditional von Neumann entropy between $A$ and $C$'s system.  This new technique provides a far more optimistic bound.  Indeed, as we will see, if the server is semi-honest, using our new key rate bound in this paper, we will see that $A$ and $B$ may distill a secret key so long as the noise in the quantum channel is less than $22.05\%$.  If the server $C$ is adversarial, then we can withstand up to $12.5\%$.  Furthermore, our new proof of security does not make any assumptions concerning the symmetry of $E$'s attack as was done in our original proof.

Thus the contributions of this paper are two-fold.  First, we improve our original security proof finding a more optimistic bound on our mediated SQKD protocol's tolerated noise level.  Our new proof in this paper also requires fewer assumptions.  Secondly, we describe an alternative proof method, which as we've already shown in \cite{SQKD-Krawec-SecurityProof}, can be applied to other semi-quantum protocols (and perhaps other QKD protocols utilizing a two-way channel).  We also show how this alternative method can produce more optimistic bounds than produced by bounding the Jensen-Shannon divergence as done in our original proof \cite{SQKD-MultiUser}.  Thus, this observation can be useful in other QKD protocol proofs and, perhaps, may lead to insight into various quantum information theoretic bounds.  Furthermore, it may be possible to adapt the techniques we use in this proof to work with other QKD protocols which utilize a two-way quantum communication channel.

\section{The Protocol}

We now review the mediated protocol of \cite{SQKD-MultiUser}.  This protocol is designed to allow two classical users $A$ and $B$ (users who can only measure and resend in the $Z$ basis or reflect qubits) to distill a secret key, known only to themselves, with the help of a quantum server $C$.  Note that, since $A$ and $B$ can only make $Z$ basis measurements, they must rely on the server to perform measurements in alternative bases.  However, neither $A$ nor $B$ trust $C$; indeed we will prove the security of this protocol assuming $C$ is adversarial.  Since this paper is concerned only with devising an improved key rate bound, we do not alter the protocol in any way.

\begin{figure}\center
\includegraphics[width=\linewidth]{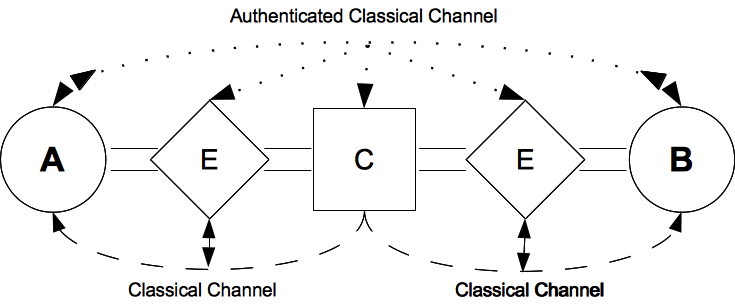}
\caption{A diagram of the scenario we consider.  Here, $A$ and $B$ are the honest, classical users who wish to establish a secret key; $E$ is a third party eavesdropper; and $C$ is the untrusted, potentially adversarial quantum server.  A quantum channel (the parallel lines in the center of the diagram) connects $C$ to $A$ and also $C$ to $B$; these channels pass through the third party eavesdropper (which may be a single entity or two separate entities).  An authenticated classical channel connects $A$ to $B$ (the dotted line); only $A$ and $B$ may write to this channel, however all parties can read from it.  Finally a classical channel connects the server $C$ to $A$ and $B$; this channel is not necessarily authenticated and the eavesdroppers may write their own messages on it.}\label{fig:diagram}
\end{figure}

This mediated protocol assumes the existence of a quantum communication channel connecting Alice ($A$) to the server $C$ and Bob ($B$) to $C$.  We do not require a quantum channel directly connecting $A$ and $B$.  Besides this, we assume an authenticated classical channel connects $A$ and $B$.  The server $C$, and indeed any other third party eavesdropper, may listen to the messages sent on this channel, but they may not send messages of their own.  Finally, we assume a classical channel connects the server to either $A$ or $B$ (or both).  See Figure \ref{fig:diagram}.  Using similar arguments as in \cite{SQKD-MultiUser}, this channel connecting $C$ to $A$ or $B$ need not be authenticated; though, indeed, better security bounds may be achieved if it is authenticated as we discuss later.  We will assume that any message sent from $C$ is received by both parties.  That is, $C$ cannot send two different messages $m_A \ne m_B$ to $A$, respectively $B$, without being caught.  Such a mechanism is easy to achieve: any message sent from $C$ to $A$ (or $B$) is then forwarded by $A$ (or $B$) to the other honest party $B$ (or $A$) using the authenticated channel.

Our protocol will utilize the Bell basis, the states of which we denote:
\begin{align*}
\ket{\Phi^+} &= \frac{1}{\sqrt{2}}(\ket{00} + \ket{11})\\
\ket{\Phi^-} &= \frac{1}{\sqrt{2}}(\ket{00} - \ket{11})\\
\ket{\Psi^+} &= \frac{1}{\sqrt{2}}(\ket{01} + \ket{10})\\
\ket{\Psi^-} &= \frac{1}{\sqrt{2}}(\ket{01} - \ket{10})
\end{align*}

We first describe the protocol assuming an honest server $C$.  The quantum communication stage of our protocol repeats the following process:

\begin{enumerate}
  \item $C$ prepares $\ket{\Phi^+}$, sending one qubit to $A$, the other to $B$.
  \item $A$ and $B$ (the classical users) choose, independently of each other, to either reflect the qubit back to $C$ or to measure and resend it.  If they measure in the $Z$ basis and resend, they save their measurement results as their potential raw key bit.  However, $A$ and $B$ do not yet reveal their choice of operation.
  \item Regardless of $A$ and $B$'s choice, $C$ will receive two qubits back from them.  $C$ then performs a Bell measurement.  If this measurement produces the result $\ket{\Phi^-}$, $C$ sends the message ``$-1$'' to both $A$ and $B$ using the classical channel.  Otherwise, for all other measurement results, $C$ sends the message ``$+1$''.
  \item $A$ and $B$ now divulge, using the authenticated classical channel, their choice in step 2 (but not their measurement results).  If $A$ and $B$ both measure and resent, and if $C$ sends the message ``$-1$'', they will \emph{accept} this iteration; that is to say, they will both use this iteration's measurement results to contribute towards their raw key.  They will not accept this iteration (i.e., it will not be used to contribute to the raw key) if $C$ sent the message ``$+1$''.
  Otherwise, if $A$ and $B$ both reflected, it should be the case that $C$ sends ``$+1$''; any other message by $C$ is counted as an error.
  All other cases are discarded.
\end{enumerate}

Assuming an honest $C$ and the lack of any channel noise or third-party eavesdropper, it is clear that the above protocol is correct (i.e., $A$ and $B$ will agree on the same raw key).  Indeed, if $A$ and $B$ both reflect, the state arriving back to $C$ on step 3 is $\ket{\Phi^+}$; thus $C$ should always respond with the message ``$+1$''.  Alternatively, if $A$ and $B$ both measure and resend, the state arriving at $C$ is $\ket{i,i}$ (where $i \in \{0,1\}$ is $A$ and $B$'s measurement result and potential raw key bit for this iteration).  $C$ will then perform a Bell basis measurement (projecting the state into one of the Bell basis states).  This measurement results in outcome $\ket{\Phi^+}$ or $\ket{\Phi^-}$ each with probability $1/2$.  Only if $C$ reports ``$-1$'' (i.e., he measures $\ket{\Phi^-}$) will $A$ and $B$ use their measurement results ``$i$'' as their raw key bit. (Note that if they also use those iterations where $C$ sends ``$+1$'', this opens a potential easy attack strategy for $C$: he can always measure in the computational basis and send ``$+1$'' each iteration.)

Let $p_M$ be the probability that $A$ or $B$ measure and resend.  Note that, without noise and assuming an honest $C$, only $p_Mp_M/2 = p_M^2/2$ iterations are expected to be accepted.  However, to improve efficiency, we may use a technique from \cite{QKD-BB84-Modification} (which was meant to improve the key rate of the BB84 protocol).  Namely, we may set $p_M$ arbitrarily close to 1.  Thus, in the asymptotic scenario, we can expect $1/2$ of the qubits sent to contribute to the raw key.  It is an open question as to whether or not a mediated SQKD protocol can be designed which improves this rate to something larger than $1/2$.

\section{Security Proof}
While this protocol's unconditional security was proven in \cite{SQKD-MultiUser}, we will now provide an alternative proof of security which, as we demonstrate later, provides a more optimistic lower bound on the protocol's key rate (we will define the key rate in the asymptotic scenario shortly).  Note that, to prove the security of this protocol, we must not only consider an adversarial $C$, but also the existence of third-party, all-powerful eavesdroppers (as shown in Figure \ref{fig:diagram}).

Our proof will work in stages: we will first prove security against an adversarial $C$, assuming $C$ is limited to collective attacks (those where he performs the same attack operation each iteration, but, unlike with an independent attack discussed earlier, he is free to postpone the measurement of his quantum memory until any future time of his choice) and assuming there are no third-party eavesdroppers.  Following this, we will prove security against general attacks (those attacks where no assumptions are made other than $C$ follows the laws of physics) and assuming the existence of third-party eavesdroppers.

\subsection{Notation}
We denote by $H(\cdot)$ the Shannon entropy function.  Given a set $\{p_1, p_2, \cdots, p_n\}$ where $\sum_ip_i = 1$ and $p_i \ge 0$, then:
\[
H(p_1, p_2, \cdots, p_n) = -\sum_{i=1}^np_i\log p_i,
\]
where all logarithms in this paper are base two unless otherwise specified.  We will occasionally use the notational shortcut $H(\{p_i\}_i)$ to mean $H(p_1, p_2, \cdots, p_n)$.  Furthermore, when $n=2$ (which forces $p_2 = 1-p_1$), we will occasionally write $h(p_1)$ to mean $H(p_1, p_2)$.

Given a density operator $\rho$, we write $S(\rho)$ to be its von Neumann entropy.  Let $\{\lambda_1, \lambda_2, \cdots, \lambda_n\}$ be the eigenvalues of finite dimensional $\rho$.  Then $S(\rho) = -\sum_i\lambda_i\log\lambda_i$.

When $\rho$ acts on a bipartite system $\mathcal{H}_A\otimes\mathcal{H}_B$, we will often write $\rho_{AB}$.  Then, later, if we write $\rho_A$ we take that to mean the result of tracing out the $\mathcal{H}_B$ portion of $\rho_{AB}$ (i.e., $\rho_A = tr_B\rho_{AB}$).  Similarly for $\rho_B$.  Similarly, for multi-partite systems (e.g., $\rho_{ABC}$ acts on a tripartite system and $\rho_{BC} = tr_A\rho_{ABC}$).

Given $\rho_{AB}$, we write $S(AB)$ to mean $S(\rho_{AB})$ and $S(B)$ to mean $S(\rho_B)$.  Finally, we write $S(A|B)$ to be the conditional von Neumann entropy defined: $S(A|B) = S(AB) - S(B) = S(\rho_{AB}) - S(\rho_B)$.

\subsection{Modeling the Protocol}
For now, we will consider only an adversarial center $C$ and not third-party eavesdroppers.  That is, the system in question consists only of $A$, $B$, and $C$ and no one else (the case when there are third-party attackers will be considered later).  We will also assume $C$ employs collective attacks.  Assuming this, a single iteration of our protocol may be described as a closed system in the Hilbert space:
\[
\mathcal{H} = \mathcal{H}_A\otimes\mathcal{H}_B\otimes\mathcal{H}_{T_A}\otimes\mathcal{H}_{T_B}\otimes\mathcal{H}_C\otimes\mathcal{H}_{cl},
\]
where:
\begin{itemize}
\item $\mathcal{H}_A$ and $\mathcal{H}_B$ are $A$ and $B$'s private registers storing their raw key bit.
\item $\mathcal{H}_{T_A}$ and $\mathcal{H}_{T_B}$ are two dimensional spaces modeling the qubit channel connecting $C$ to $A$ and $C$ to $B$ respectively (they are the \emph{transit} space).
\item $\mathcal{H}_C$ is $C$'s private quantum system.
\item $\mathcal{H}_{cl}$ is a two-dimensional subspace, spanned by orthonormal basis $\{\ket{+1}, \ket{-1}\}$ used to model the classical message $C$ sends on this iteration.
\end{itemize}

We model the protocol in the same way as in \cite{SQKD-MultiUser}, the details of which we now quickly review.  At start, $C$, who we now assume is fully adversarial, prepares, not necessarily $\ket{\Phi^+}$ as prescribed by the protocol, but instead the state $\ket{\phi_0} = \sum_{i,j}\alpha_{i,j}\ket{i,j}_{T_A,T_B}\otimes\ket{c_{i,j}}$ where the $\ket{c_{i,j}}$ are arbitrary normalized, though not necessarily orthogonal, states in $\mathcal{H}_C$.

As shown in our original proof, there is no advantage to $C$ in sending the above state, versus sending the far simpler state: $\ket{{\psi}_0} = \sum_{i,j}\alpha_{i,j}\ket{i,j}_{T_A,T_B}$.  That is, there is no advantage to $C$ in preparing a state, on step (1) of the protocol, that is entangled with his private quantum memory.  The proof of this, which may be found in \cite{SQKD-MultiUser}, uses a technique introduced in \cite{SQKD-Single-Security}.  Thus, when analyzing the security of this protocol, we may assume the state $C$ sends is unentangled with $\mathcal{H}_C$.

Following $A$ and $B$'s operation (either measuring and resending, or reflecting), the qubits return to $C$.  The server, $C$, is now permitted to perform any operation of his choice, potentially entangling the qubits with his private quantum memory.  Since he must also send a single classical bit to both $A$ and $B$ (from our earlier discussion, it is impossible for $C$ to send different messages to $A$ and $B$ thus we may assume he sends only a single message and it is received by both), his attack is modeled as a \emph{quantum instrument} \cite{QC-Instrument} $\mathcal{I}$ which acts on density operator $\rho = \rho_{T_AT_BC}$ as follows:
\begin{equation}\label{eq:instrument}
\mathcal{I}(\rho) = \ket{+1}\bra{+1}_{cl}\otimes\sum_{i=1}^{N_0}E_{i,0}\rho E_{i,0}^* + \ket{-1}\bra{-1}_{cl}\otimes\sum_{i=1}^{N_1}E_{i,1}\rho E_{i,1}^*,
\end{equation}
where the $E_{i,j}$ satisfy:
\[
\sum_{i=1}^{N_0} E_{i,0}^*E_{i,0} + \sum_{i=1}^{N_1}E_{i,1}^*E_{i,1} = I.
\]
(Above, $I$ is the identity operator.)

We may assume, without loss of generality that $N_0$ and $N_1$ are both finite.  Finally, it can be shown (see \cite{SQKD-MultiUser} for details) that this attack may be represented, without loss of power to $C$, by a unitary operator $U = U_\mathcal{I}$ acting on a larger, though still finite dimensional, Hilbert space.  Providing $C$ with this larger space potentially increases his power, thus providing us with a lower bound on the security of our protocol.  Working with unitary attack operators turns out to be far simpler than working with attacks of the form in Equation \ref{eq:instrument}.

This unitary attack operator, $U$, acts on $\mathcal{H}_{T_A}\otimes\mathcal{H}_{T_B}\otimes\mathcal{H}_C\otimes\mathcal{H}_{cl}$, where $\mathcal{H}_C$ has been suitably expanded.  After $A$ and $B$'s operation, $C$ will apply unitary $U$.  $C$ will then perform a projective measurement on the $\mathcal{H}_{cl}$ system in the $\{\ket{+1}, \ket{-1}\}$ basis.  This measurement determines the message he sends.  The post-measurement state represents the state of his ancilla in the event he sends that particular message.  Such a procedure is mathematically equivalent to his use of $\mathcal{I}$ as proven in \cite{SQKD-MultiUser}.

Clearly, a single iteration of the protocol, conditioning on the event the iteration is accepted (i.e., $A$ and $B$ both measure and resend, and $C$ sends ``$-1$'') may be described by a density operator of the form:
\[
\rho_{ABC} = \sum_{i,j}p_{i,j}\ketbra{i,j}_{AB}\otimes\sigma_C^{(i,j)},
\]
where $p_{i,j}$ is the probability that $A$ and $B$'s raw key bit is $i$ and $j$ respectively and $\sigma_C^{(i,j)}$ is the state of $C$'s quantum memory in that event.

Following $N$ iterations of the quantum communication stage, assuming collective attacks, the overall system is in the state $\rho_{ABC}^{\otimes N}$.  $A$ and $B$ will then perform parameter estimation, error correction, and privacy amplification protocols (see \cite{QKD-survey} for more information on these, now standard, processes) resulting in a secret key of size $\ell(N) \le N$ (possibly $\ell(N) = 0$ if $C$ has too much information on the raw key).  We are interested in the ratio of secret key bits to raw key bits as the latter approaches infinity; that is, we are interested in the \emph{key rate}, denoted $r$, in the \emph{asymptotic scenario} \cite{QKD-survey}:
\[
r = \lim_{N \rightarrow \infty} \frac{\ell(N)}{N}.
\]

For a state of the form $\rho^{\otimes N}$, it was shown in \cite{QKD-Winter-Keyrate,QKD-renner-keyrate,QKD-renner-keyrate2} that the key rate is:
\begin{equation}
r = \inf(I(A:B) - I(A:C)) = \inf(S(A|C) - S(A|B)).
\end{equation}
Here $I(A:B)$ is the (classical) mutual information held between $A$ and $B$'s system; $I(A:C)$ is the quantum mutual information between $A$ and $C$; the conditional entropy $S(A|C)$ was defined earlier in the notation section; $H(A|B)$ is the conditional Shannon entropy defined: $H(A|B) = H(AB) - H(B)$.  Finally, the infimum is over all attack operators which induce the observed statistics (e.g., the observed error rate).

Note that the first equation, involving mutual information, is due to \cite{QKD-Winter-Keyrate}, the second, equivalent version, is from \cite{QKD-renner-keyrate,QKD-renner-keyrate2} (showing the two are equal is trivial).  While in our previous paper \cite{SQKD-MultiUser}, we used the key rate equation based on mutual information, in this paper we will bound $r$ by bounding the conditional entropy.  Though these two equations produce equal results, the technique we use to bound the latter, as we show in this paper, allows us to provide a more optimistic lower bound on $r$.

\subsection{The New Key Rate Bound}
We must now describe the state of $A$, $B$, and $C$'s system after a single iteration of the protocol, conditioning on the event that this iteration is used to contribute towards the raw key (i.e., both $A$ and $B$ measure and resend, and $C$ sends ``$-1$'').  Let $\ket{\psi_0} = \sum_{i,j}\alpha_{i,j}\ket{i,j}_{T_A,T_B}$ be the initial state prepared and sent by $C$ (from our earlier discussion this is without loss of generality), where the qubit $T_A$ is sent to $A$ and the qubit $T_B$ is sent to $B$.  Now, assume both $A$ and $B$ measure and resend (other cases, though potentially useful for parameter estimation, do not contribute towards the raw key and thus are not considered for the time being).  After $A$ and $B$'s measure and resend operation, the state is clearly:

\[
\rho_1 = \sum_{i,j\in\{0,1\}} |\alpha_{i,j}|^2\ketbra{i,j}_{A,B}\otimes\ketbra{i,j}_{T_A,T_B}.
\]

At this point, the transit system returns to $C$'s control where he will perform an arbitrary, without loss of generality unitary, operator acting on $\mathcal{H}_{T_A}\otimes\mathcal{H}_{T_B}\otimes\mathcal{H}_C\otimes\mathcal{H}_{cl}$ (since we are working with collective attacks, the latter two subspaces are assumed to be cleared to some ``zero'' state).  We write $U$'s action on basis states as follows:

\begin{equation}\label{eq:U-action}
\ket{i,j} \overset{U}\mapsto \ket{+1, e_{i,j}} + \ket{-1, f_{i,j}},
\end{equation}
where $\ket{e_{i,j}}$ and $\ket{f_{i,j}}$ are arbitrary, not necessarily normalized nor orthogonal, states in $\mathcal{H}_{T_A}\otimes\mathcal{H}_{T_B} \otimes \mathcal{H}_C$.  Unitarity of $U$ imposes various restrictions on these states which will become important later in our analysis.  Note that, unlike our analysis in our original paper, we do not make any symmetry assumptions at this point.

Following $C$'s operation, and conditioning on the event that $C$ sends ``$-1$'' (if $C$ sends ``$+1$'' the iteration is discarded and thus does not contribute to the raw key - though such iterations will be important later for parameter estimation as we discuss shortly), the final state is:

\begin{equation}\label{eq:final-state-ABC}
\rho_{ABC} = \frac{1}{p_a} \sum_{i,j}|\alpha_{i,j}|^2 \ketbra{i,j}_{AB} \otimes \ketbra{f_{i,j}}_C,
\end{equation}
where:
\begin{equation}\label{eq:pa}
p_a = \sum_{i,j}|\alpha_{i,j}|^2 \braket{f_{i,j}|f_{i,j}}.
\end{equation}
(Note that we have disregarded the $\mathcal{H}_{cl}$ portion of the above state as it is projected to $\ketbra{-1}_{cl}$ and thus not needed; we have also abused notation slightly by ``absorbing'' the $\mathcal{H}_{T_A}\otimes\mathcal{H}_{T_B}$ subspaces into $\mathcal{H}_C$.

Our goal is to compute a bound on $S(A|C) - H(A|B)$.  We will first bound $S(A|C)$.  However, the high dimensionality of the system proves a hinderance.  To overcome this, we will employ a technique first proposed in \cite{QKD-keyrate-general}, and later adapted successfully by us, to the key rate computation of a different (not a mediated) SQKD protocol in \cite{SQKD-Krawec-SecurityProof}.  This technique requires us to condition on a new random variable of our choice.  By appending a carefully chosen auxiliary system, we can simplify the entropy computations.  Due to the strong sub additivity of von Neumann entropy, it holds that, for any tripartite system $\mathcal{H}_{A}\otimes\mathcal{H}_{C}\otimes\mathcal{H}_X$, we have:
\[
S(A|C) \ge S(A|CX) \Longrightarrow S(A|C) - H(A|B) \ge S(A|CX) - H(A|B),
\]
thus allowing us to find a lower bound on the key rate of this mediated SQKD protocol.  Appending this system $\mathcal{H}_X$ and providing $C$ access to it, though unrealistic, does allow us to compute a lower bound on the key rate; the realistic case, then, where $C$ does not have this additional information $\mathcal{H}_X$ can only be better for $A$ and $B$.

The system we append, denoted $\mathcal{H}_X$, is two-dimensional and spanned by orthonormal basis states $\{\ket{C}, \ket{W}\}$ where $\ket{C}$ will be used to denote the event that $A$ and $B$'s raw key bit is correct, while $\ket{W}$ will be used to describe the event that their raw key bits are wrong.

Incorporating this system into $\rho_{ABC}$ yields the state:
\begin{align}
\rho_{ABCX} &= &&\frac{1}{p_a}\ketbra{C}\otimes\left(|\alpha_{0,0}|^2\ketbra{0,0}_{AB}\otimes\ketbra{f_{0,0}}\right.\label{eq:final-state-ABCX}\\
&&&\left.+ |\alpha_{1,1}|^2\ketbra{1,1}_{AB}\otimes\ketbra{f_{1,1}}\right) \notag\\ \notag\\
&+&&\frac{1}{p_a}\ketbra{W}\otimes\left(|\alpha_{0,1}|^2\ketbra{0,1}_{AB}\otimes\ketbra{f_{0,1}}\right.\notag\\
&&&\left.+|\alpha_{1,0}|^2\ketbra{1,0}_{AB}\otimes\ketbra{f_{1,0}}\right).\notag
\end{align}

Given such a state, we may more readily compute $S(A|CX) = S(ACX) - S(CX)$.  Indeed, it is not difficult to see that:
\begin{equation}\label{eq:entropy-ACX}
S(ACX) = H\left( \left\{ \frac{1}{p_a}|\alpha_{i,j}|^2 \braket{f_{i,j}|f_{i,j}}\right\}_{i,j}\right).
\end{equation}
(Choosing a suitable basis, we may write $\rho_{ACX}$ - which is the result of tracing out $B$ from Equation \ref{eq:final-state-ABCX} - as a diagonal matrix with diagonal elements equal to $\frac{1}{p_a} |\alpha_{i,j}|^2\braket{f_{i,j}|f_{i,j}}$.)

We must now compute an upper bound on $S(CX)$ which will provide us with a lower bound on the key rate equation.  Indeed, if $S(CX) \le \eta$ then $S(A|C) \ge S(A|CX) \ge S(ACX) - \eta$.

Define the following:
\begin{align}
p_C &= \frac{1}{p_a}(|\alpha_{0,0}|^2\braket{f_{0,0}|f_{0,0}} + |\alpha_{1,1}|^2\braket{f_{1,1}|f_{1,1}})\label{eq:pc}\\ \notag\\
p_W &= \frac{1}{p_a}(|\alpha_{0,1}|^2\braket{f_{0,1}|f_{0,1}} + |\alpha_{1,0}|^2\braket{f_{1,0}|f_{1,0}})\label{eq:pw}.
\end{align}
Clearly, $p_C$ is the probability that $A$ and $B$'s raw key bit is correct (they match) while $p_W$ is the probability that they are wrong.  For the following, assume that both $p_C$ and $p_W$ are both strictly positive.  The case when $p_W$ is zero is similar as we will comment later.  Of course if $p_C = 0$ then there is far too much noise and $A$ and $B$ should abort. (If $p_C = 0$ all their key bits are wrong!)  Note that $p_C$ and $p_W$ are parameters that may be estimated by $A$ and $B$.  Also note that $A$ and $B$ may estimate the quantities $\braket{f_{i,j}|f_{i,j}}$ which are simply the probabilities that $C$ sends ``$-1$'' in the event $A$ and $B$ both measure and resend $\ket{i,j}$.

Tracing out $A$ and $B$ from Equation \ref{eq:final-state-ABCX} yields:
\begin{align}
\rho_{CX} &= \frac{1}{p_a}\ketbra{C}\otimes\left( |\alpha_{0,0}|^2 \ketbra{f_{0,0}} + |\alpha_{1,1}|^2 \ketbra{f_{1,1}} \right)\label{eq:final-state-CX}\\
&+\frac{1}{p_a}\ketbra{W} \otimes \left( |\alpha_{0,1}|^2 \ketbra{f_{0,1}} + |\alpha_{1,0}|^2 \ketbra{f_{1,0}} \right).\notag \\ \notag\\
&= p_C\ketbra{C} \otimes \sigma_C + p_W\ketbra{W} \otimes \sigma_W,\notag
\end{align}
where $\sigma_x$ are the following (unit trace - recall we are assuming for now that both $p_C$ and $p_W$ are non-zero) density operators:
\begin{align}
\sigma_C &= \frac{|\alpha_{0,0}|^2 \ketbra{f_{0,0}} + |\alpha_{1,1}|^2\ketbra{f_{1,1}}}{|\alpha_{0,0}|^2\braket{f_{0,0}|f_{0,0}} + |\alpha_{1,1}|^2\braket{f_{1,1}|f_{1,1}}}.\\ \notag\\
\sigma_W &= \frac{|\alpha_{0,1}|^2 \ketbra{f_{0,1}} + |\alpha_{1,0}|^2\ketbra{f_{1,0}}}{|\alpha_{0,1}|^2\braket{f_{0,1}|f_{0,1}} + |\alpha_{1,0}|^2\braket{f_{1,0}|f_{1,0}}}.
\end{align}

It is not difficult to show (see, for instance, \cite{SQKD-Krawec-SecurityProof} for a proof) that the entropy of such a system is simply:
\begin{align}
S(CX) &= H(p_C, p_W) + p_WS(\sigma_W) + p_CS(\sigma_C)\notag\\
&\le H(p_C, p_W) + p_W + p_CS(\sigma_C).\label{eq:entropy-bound}
\end{align}
The inequality above follows from the fact that $S(\sigma_W) \le \log \dim \sigma_W$.  Since $\sigma_C$ and $\sigma_W$ are both two-dimensional, $S(\sigma_W) \le 1$.  Note that it is not difficult to show that if $p_W = 0$, then $|\alpha_{0,1}|^2\ketbra{f_{0,1}} + |\alpha_{1,0}|^2\ketbra{f_{1,0}} \equiv 0$.  Thus, this term does not appear in Equation \ref{eq:final-state-CX}, and so the bound in Equation \ref{eq:entropy-bound} holds even in this case.

Obviously if the noise is small, than $p_W$ should also be small.  All that remains, therefore, is to upper bound $S(\sigma_C)$.

We may write, without loss of generality, $\ket{f_{0,0}} = x\ket{f}$ and $\ket{f_{1,1}} = y\ket{f} + z\ket{\zeta}$, where $\braket{f|f} = \braket{\zeta|\zeta} = 1$, $\braket{f|\zeta} = 0$, and $x,y,z \in \mathbb{C}$.  This of course implies:
\begin{align}
&|x|^2 = \braket{f_{0,0}|f_{0,0}}\label{eq:identity-ev-1}\\
&|y|^2+|z|^2 = \braket{f_{1,1}|f_{1,1}}\label{eq:identity-ev-2}\\
&x^*y = \braket{f_{0,0}|f_{1,1}} \Longrightarrow |y|^2 ={|\braket{f_{0,0}|f_{1,1}}|^2}/{|x|^2}.\label{eq:identity-ev-3}
\end{align}

Using this $\{\ket{f}, \ket{\zeta}\}$ basis, we may write $\sigma_C$ as:

\[
\sigma_C = q_0 \left( \begin{array}{ccc}
|\alpha_{0,0}|^2 |x|^2 + |\alpha_{1,1}|^2|y|^2 &,& |\alpha_{1,1}|^2y^*z \\\\
|\alpha_{1,1}|^2yz^* &,& |\alpha_{1,1}|^2|z|^2\end{array}\right),
\]
where:
\[
q_0 = (|\alpha_{0,0}|^2\braket{f_{0,0}|f_{0,0}} + |\alpha_{1,1}|^2\braket{f_{1,1}|f_{1,1}})^{-1} = [|\alpha_{0,0}|^2|x|^2 + |\alpha_{1,1}|^2(|y|^2+|z|^2)]^{-1}.
\]

The eigenvalues of this matrix are easily computed to be:
\begin{align*}
\lambda_{\pm} &= \frac{1}{2} \pm \frac{q_0}{2}\sqrt{\left(|\alpha_{0,0}|^2|x|^2 + |\alpha_{1,1}|^2|y|^2 - |\alpha_{1,1}|^2|z|^2\right)^2 + 4 |\alpha_{1,1}|^4|y|^2|z|^2}\\
&=\frac{1}{2} \pm \frac{q_0}{2}\sqrt{ \left( |\alpha_{0,0}|^2F_{0,0} + |\alpha_{1,1}|^2[2|y|^2 - F_{1,1}] \right)^2 + 4 |\alpha_{1,1}|^4|y|^2(F_{1,1} - |y|^2)},
\end{align*}
where above we have defined $F_{i,i} = \braket{f_{i,i}|f_{i,i}}$ and have used the identities \ref{eq:identity-ev-1} and \ref{eq:identity-ev-2}.  Now, define $\Delta = |\alpha_{0,0}|^2F_{0,0} - |\alpha_{1,1}|^2F_{1,1}$ and, continuing, we have:
\begin{align*}
\lambda_{\pm} &= \frac{1}{2} \pm \frac{q_0}{2}\sqrt{\left(\Delta + 2 |\alpha_{1,1}|^2|y|^2\right)^2 + 4|\alpha_{1,1}|^4|y|^2F_{1,1} - 4|\alpha_{1,1}|^4|y|^4}\\
&= \frac{1}{2} \pm \frac{q_0}{2}\sqrt{\Delta^2 + 4|\alpha_{1,1}|^2|y|^2\left(\Delta + |\alpha_{1,1}|^2F_{1,1}\right)}\\
&= \frac{1}{2} \pm \frac{q_0}{2}\sqrt{\Delta^2 + 4|\alpha_{1,1}|^2|y|^2|\alpha_{0,0}|^2|x|^2}.
\end{align*}
Finally, using identity \ref{eq:identity-ev-3} (and that $|x|^2 = F_{0,0}$), we have:
\begin{equation}
\lambda_{\pm} = \frac{1}{2} \pm \frac{q_0}{2}\sqrt{\Delta^2 + 4|\alpha_{0,0}|^2|\alpha_{1,1}|^2|\braket{f_{0,0}|f_{1,1}}|^2}
\end{equation}

Note that $\Delta, q_0, |\alpha_{0,0}|^2$, and $|\alpha_{1,1}|^2$ are all parameters that $A$ and $B$ may estimate.  Computing a bound on $|\braket{f_{0,0}|f_{1,1}}|^2$ may be achieved using the error rate when both $A$ and $B$ reflect as we demonstrate later.

%Indeed, let $Q_R$ be the probability that $C$ sends ``$-1$'' if both $A$ and $B$ reflect (this is an error).  From the linearity of $C$'s attack operator $U$, we have:
%\[
%U\ket{\psi_0} = \sum_{i,j}\alpha_{i,j}U\ket{i,j} = \ket{+1}\left(\sum\alpha_{i,j}\ket{e_{i,j}}\right) + \ket{-1}\left(\sum_{i,j}\alpha_{i,j}\ket{f_{i,j}}\right)
%\]
%and so:
%\begin{equation}\label{eq:Qr}
%Q_R = \left| \sum_{i,j}\alpha_{i,j}\ket{f_{i,j}}\right|^2.
%\end{equation}
%This expression, as we demonstrate later, may be used to bound the quantity $|\braket{f_{0,0}|f_{1,1}}|^2$.

Combining everything together yields the bound:
\begin{align*}
S(A|C) \ge H\left( \left\{ \frac{1}{p_a}|\alpha_{i,j}|^2 \braket{f_{i,j}|f_{i,j}}\right\}_{i,j}\right) - H(p_C, p_W) - p_W - p_CH(\lambda_+, \lambda_-).
\end{align*}

Note that $H(\lambda_+, \lambda_-) = h(\lambda_+)$ takes its maximum when $\lambda_+ = \frac{1}{2}$.  It is not difficult to see that, as $|\braket{f_{0,0}|f_{1,1}}|^2 \ge 0$ increases, $\lambda_+$ increases.  Furthermore, when $|\braket{f_{0,0}|f_{1,1}}|^2 = 0$, $\lambda_+ \ge \frac{1}{2}$.  Thus, as $|\braket{f_{0,0}|f_{1,1}}|^2$ increases, $h(\lambda_+)$ necessarily decreases.  By finding a lower-bound $\mathcal{F}$ (which will be a function of certain observed parameters as we will soon discuss) such that $|\braket{f_{0,0}|f_{1,1}}|^2 \ge \mathcal{F}$, we will have an upper-bound on $h(\lambda_+)$ and thus a lower bound on $S(A|CX)$.  That is to say, if we define:
\begin{equation}
\tilde{\lambda} = \frac{1}{2} + \frac{q_0}{2}\sqrt{\Delta^2 + 4|\alpha_{0,0}|^2|\alpha_{1,1}|^2\mathcal{F}},
\end{equation}
then:
\[
\frac{1}{2} \le \tilde{\lambda} \le \lambda_+ \Longrightarrow h(\lambda_+) \le h(\tilde{\lambda}),
\]
%as can be seen more clearly in Figure \ref{fig:entropy-bound}.
This implies:
\begin{equation}\label{eq:entropy-bound-final}
S(A|C) \ge H\left( \left\{ \frac{1}{p_a}|\alpha_{i,j}|^2 \braket{f_{i,j}|f_{i,j}}\right\}_{i,j}\right) - H(p_C, p_W) - p_W - p_Ch(\tilde{\lambda}).
\end{equation}

This concludes our bound on $S(A|C)$.  Determining a bound on $\mathcal{F}$ (needed to bound $h(\lambda_+)$) can be easily determined from the probability that $C$ sends ``$-1$'' if both $A$ and $B$ reflect (note that $C$ should send ``$+1$'' if they both reflect - $C$ sending ``$-1$'' in this event is counted as an error).  In the next section, we will demonstrate this in two specific cases comparing our new result with our older one from \cite{SQKD-MultiUser}.

Of course, computing $H(A|B)$, the last remaining term from the key rate equation, is easy after parameter estimation.  Indeed, let $p(a,b)$ be the probability of $A$'s raw key bit being $a$ and $B$'s raw key bit being $b$.  From Equation \ref{eq:final-state-ABC}, we see these values are:
\begin{equation}\label{eq:pab}
p(i,j) = \frac{1}{p_a}|\alpha_{i,j}|^2 \braket{f_{i,j}|f_{i,j}}
\end{equation}
Also, if we let $p(b)$ be the probability that $B$'s raw key bit is $b$ (i.e., $p(b) = p(0,b) + p(1,b)$) then our final key rate bound is found to be:
\begin{align*}
r &\ge H\left( \left\{ \frac{1}{p_a}|\alpha_{i,j}|^2 \braket{f_{i,j}|f_{i,j}}\right\}_{i,j}\right) - H(p_C, p_W) - p_W - p_Ch(\tilde{\lambda})\\
& - H\left(\{p(a,b)\}_{a,b}\right) + h(p(0)).
\end{align*}
Using Equation \ref{eq:pab}, this can be simplified to:
\begin{equation}\label{eq:key-rate-bound}
r \ge h(p(0)) - H(p_C, p_W) - p_W - p_Ch(\tilde{\lambda})
\end{equation}

Note that all terms in the above expression, with the exception of $\tilde{\lambda}$, are directly observable by $A$ and $B$.  Indeed, $|\alpha_{i,j}|^2$ is simply the probability that $A$ and $B$ measure $\ket{i,j}$.  The quantity $\braket{f_{i,j}|f_{i,j}}$ is the probability that $C$ sends ``$-1$'' in the event $A$ and $B$ measure and resend $\ket{i,j}$.  If $A$ and $B$ divulge complete information about certain randomly chosen iterations (in particular their choices and measurement results if applicable), this information is easily estimated.  The only parameter that cannot be directly observed is $\tilde{\lambda}$; in particular the quantity $|\braket{f_{0,0}|f_{1,1}}|^2$ upon which that eigenvalue depends.  They may, however, estimate it using the probability that $C$ sends ``$-1$'' if $A$ and $B$ both reflect (the probability of which $A$ and $B$ may estimate).  We will show how this is done in the next section when we actually evaluate the above key rate expression for certain scenarios.

\subsection{General Attacks and Third Party Eavesdroppers}
We considered only collective attacks above.  However, if $A$ and $B$ permute their raw key bits, using a randomly chosen (and publicly disclosed) permutation, the protocol becomes permutation invariant \cite{QKD-renner-keyrate}.  Thus the results in \cite{QKD-general-attack,QKD-general-attack2} apply showing that, to prove security against general attacks (where $C$ is allowed to perform any operation of his choice - perhaps altering his attack operator each iteration) it is sufficient to prove security against collective attacks.  Furthermore, since we are considering the asymptotic scenario in this paper, our key rate bound is equivalent in both cases.

Finally, it is clear that any attack by a third-party eavesdropper (including attacks whereby the eavesdropper alters $C$'s classical messages - recall that channel is not authenticated) can simply be ``absorbed'' into $C$'s attack operator $U$.  Thus our bound holds even in this case.

\section{Evaluation}
Our key rate bound applies in the most general of cases.  $A$ and $B$ must simply observe certain parameters and, based on these, they may determine a lower bound on their secret key fraction (bounding $|\braket{f_{0,0}|f_{1,1}}|^2$ may also be achieved using these parameters as we discuss shortly).  Of course, due to its reliance on many parameters, it is difficult to visualize this bound here in this paper.  However, we may consider certain scenarios which $A$ and $B$ may encounter and evaluate our bound based on these particular scenarios.  First, we will consider the case of a semi-honest server and a noisy quantum channel.  Later we will consider an adversarial server whose attack is ``symmetric'' (a common assumption in QKD security proofs).  These two scenarios were considered in our earlier work \cite{SQKD-MultiUser} and so will allow us to compare our new results with our old showing the superiority of our new bound in both of these scenarios.

\subsection{Bounding $|\braket{f_{0,0}|f_{1,1}}|^2$}
We now show how to bound the quantity $|\braket{f_{0,0}|f_{1,1}}|^2$ based on the probability that $C$ sends ``$-1$'' if $A$ and $B$ had sent the Bell state $\ket{\Phi^+}$ to $C$.  Obviously this is not a parameter that is directly observable ($A$ and $B$ are classical and so cannot prepare $\ket{\Phi^+}$).  However, it may be bounded, as we show later, using the error rate in those iterations where $A$ and $B$ reflected (i.e., the probability that $C$ sends ``$-1$'' if $A$ and $B$ both reflect).

Consider the action of $C$'s attack operator $U$ on Bell basis states.  This is:
\begin{align*}
U\ket{\Phi^+} &= \ket{+1, g_0} + \ket{-1, h_0}\\
U\ket{\Phi^-} &= \ket{+1, g_1} + \ket{-1, h_1}\\
U\ket{\Psi^+} &= \ket{+1, g_2} + \ket{-1, h_2}\\
U\ket{\Psi^-} &= \ket{+1, g_3} + \ket{-1, h_3},
\end{align*}
where each $\ket{g_i}$ and $\ket{h_i}$ are linear functions of $\ket{e_{k,l}}$ and $\ket{f_{k,l}}$ respectively (see Equation \ref{eq:U-action}).  In particular, we have $\ket{h_0} = \frac{1}{\sqrt{2}}(\ket{f_{0,0}} + \ket{f_{1,1}})$.  Imagine, for the time, that, in addition to the other parameters mentioned in the last section, $A$ and $B$ are also able to estimate the parameter $\braket{h_0|h_0}$:
\begin{equation}\label{eq:h0}
\braket{h_0|h_0} = \frac{1}{2}(\braket{f_{0,0}|f_{0,0}} + \braket{f_{1,1}|f_{1,1}} + 2Re\braket{f_{0,0}|f_{1,1}}).
\end{equation}
This then provides an estimate of $|\braket{f_{0,0}|f_{1,1}}|^2$.  Indeed, if $Re\braket{f_{0,0}|f_{1,1}} = x$, then of course $|\braket{f_{0,0}|f_{1,1}}|^2 = Re^2\braket{f_{0,0}|f_{1,1}} + Im^2\braket{f_{0,0}|f_{1,1}} \ge Re^2\braket{f_{0,0}|f_{1,1}} = |x|^2$.  Thus:
\begin{align}
\braket{h_0|h_0} &= \eta\notag\\
\Rightarrow |\braket{f_{0,0}|f_{1,1}}|^2 &\ge \left(\frac{1}{2}\braket{f_{0,0}|f_{0,0}} + \frac{1}{2}\braket{f_{1,1}|f_{1,1}} - \eta\right)^2\label{eq:bound-equal-f00f11}.
\end{align}

This quantity $\braket{h_0|h_0}$ is simply the probability that $C$ sends ``$-1$'' if $A$ and $B$ jointly send the state $\ket{\Phi^+}$ (this should be small).  Obviously they cannot estimate this parameter directly in reality (they are neither of them quantum).  However, as we will show later, they are able to compute an upper bound on it.  In particular, if they bound $\braket{h_0|h_0} \le \eta$, then, for $\eta$ small enough (and recall it should be small since $C$ is supposed to send ``$+1$'' if he receives $\ket{\Phi^+}$), it holds that:
\begin{align}
\braket{h_0|h_0} &\le \eta\notag\\
\Rightarrow Re\braket{f_{0,0}|f_{1,1}} &\le \eta - \frac{1}{2}\braket{f_{0,0}|f_{0,0}} - \frac{1}{2}\braket{f_{1,1}|f_{1,1}} \le 0\notag\\
\Rightarrow Re^2\braket{f_{0,0}|f_{1,1}} &\ge \left(\frac{1}{2}\braket{f_{0,0}|f_{0,0}} + \frac{1}{2}\braket{f_{1,1}|f_{1,1}} - \eta\right)^2\notag\\
\Rightarrow |\braket{f_{0,0}|f_{1,1}}|^2 &\ge \left(\frac{1}{2}\braket{f_{0,0}|f_{0,0}} + \frac{1}{2}\braket{f_{1,1}|f_{1,1}} - \eta\right)^2.\label{eq:bound-general-f00f11}
\end{align}
(We see that $\eta$ must be small enough so that the right hand side of the second inequality is negative.)

Both of these two identities will be useful in the subsequent examples where we bound $\eta$ as a function of $Q$ - the probability that $A$ and $B$'s measurement results differ - and $p_w$ - the probability that $C$ sends the wrong message when $A$ and $B$ both reflect (i.e., he sends ``$-1$''); with $Q = p_w = 0$ implying $\eta = 0$.

\subsection{Semi-Honest Center}

In this first example, we will consider the case where $C$ is \emph{semi-honest} - that is, he always follows the protocol exactly, preparing the correct state in step one of the protocol, performing the correct measurement in step three, and reporting, honestly, his measurement result.  Beyond that, however, he is free to do whatever he likes, for instance, he can listen in on the public communication channel to try and gain additional information on the key.  This is not only a practically relevant scenario to analyze, but it also allows us to compare our new key rate bound with our old from \cite{SQKD-MultiUser}.

Besides the semi-honest sever, we will also assume a noisy quantum channeled modeled using two independent depolarization channels, one for the forward direction (qubits traveling from $C$ to $A$ and $B$) with parameter $p$, the other for the reverse (qubits returning from $A$ and $B$ to $C$) with parameter $q$.  That is, if the joint qubit state is $\rho$ (a density operator acting on a four dimensional Hilbert space), then the depolarization channel with parameter $p$ is:
\[
\mathcal{E}_p(\rho) = (1-p)\rho + \frac{p}{4}I,
\]
where $I$ is the identity operator.  Furthermore, for the remainder of this sub-section, we will relabel the Bell basis states as follows: $\ket{\Phi^+} = \ket{\phi_0}$, $\ket{\Phi^-} = \ket{\phi_1}$, $\ket{\Psi^+} = \ket{\phi_2}$, and $\ket{\Psi^-} = \ket{\phi_3}$.

To compute a key rate bound, we need only compute a few parameters.  First are the probabilities that $C$ sends ``$-1$'' if $A$ and $B$ measure and resend the value $\ket{i,j}$ (i.e., we estimate the value $\braket{f_{i,j}|f_{i,j}}$).  These are easy to compute given our assumptions of a semi-honest server and depolarization channels.  For instance, if $A$ and $B$ measure and resend $\ket{0,0}$, then the state arriving at $C$'s lab is:
\[
\mathcal{E}_q(\ketbra{0,0}) = (1-q)\ketbra{0,0} + \frac{q}{4}\sum_{i,j\in\{0,1\}}\ketbra{i,j}.
\]

$C$ now performs a Bell measurement of this system, with the message ``$-1$'' being sent only if he measures $\ketbra{\phi_1}$ (i.e., $\ketbra{\Phi^-}$).  It is easy to see that the probability of him sending this message, given state $\mathcal{E}_q(\ketbra{0,0})$ is:
\[
\frac{1-q}{2} + \frac{q}{4} = \frac{2-q}{4}.
\]
Thus $\braket{f_{0,0}|f_{0,0}} = (2-q)/4$.  Using a similar process, the remaining $\braket{f_{i,j}|f_{i,j}}$ may be computed:
\begin{align}
\braket{f_{0,0}|f_{0,0}} &= \frac{2-q}{4} = \braket{f_{1,1}|f_{1,1}}\label{eq:dc:f00}\\
\braket{f_{0,1}|f_{0,1}} &= \frac{q}{4} = \braket{f_{1,0}|f_{1,0}}.\label{eq:dc:f01}
\end{align}

Next, we compute $|\alpha_{i,j}|^2$ (the probability that $A$ and $B$ measure $\ket{i,j}$).  Since we are assuming $C$ is semi-honest and thus he prepared $\ket{\phi_0}$ initially, these are computed using the state:
\[
\mathcal{E}_p(\ketbra{\phi_0}) = (1-p)\ketbra{\phi_0} + \frac{p}{4}\sum_{i=0}^3\ketbra{\phi_i}.
\]
Clearly:
\begin{align*}
|\alpha_{0,0}|^2 &= \frac{2-p}{4} = |\alpha_{1,1}|^2\\
|\alpha_{0,1}|^2 &= \frac{p}{4} = |\alpha_{1,0}|^2.
\end{align*}

Thus, the value of $p$ may be estimated by $A$ and $B$ using the observable parameters $|\alpha_{i,j}|^2$, while $q$ can be estimated using the observable parameters $\braket{f_{i,j}|f_{i,j}}$.  Furthermore, we have that $Q$, which is the probability that $A$ and $B$'s measurement results are different, is $Q = |\alpha_{0,1}|^2 + |\alpha_{1,0}|^2 = p/2$.

$A$ and $B$ are now able to compute $p_a$, $p_C$, and $p_W$ in terms of $p$ and $q$.  All that remains is to bound $|\braket{f_{0,0}|f_{1,1}}|^2$.  We will use our above discussion and instead determine a value $\eta$ such that $\braket{h_0|h_0}$ (which is the probability that $C$ sends ``$-1$'' if the joint state leaving $A$ and $B$'s lab is $\ket{\phi_0}$) is upper bounded by $\eta$.

In the general case, which we consider next, this value $\eta$ can only be estimated (since $A$ and $B$ are classical users, they cannot prepare a Bell state to directly observe $\braket{h_0|h_0}$).  However, given the assumptions in this sub-section, they may in fact compute a value for $\eta$ based on $q$.  Indeed, if the state leaving $A$ and $B$ is $\ket{\phi_0}$, then the system, when it reaches $C$, has evolved via:
\[
\mathcal{E}_q(\ketbra{\phi_0}) = (1-q)\ketbra{\phi_0} + \frac{q}{4}\sum_{i=0}^3\ketbra{\phi_i}.
\]
Since $C$ is semi-honest, he will only send ``$-1$'' if he measures $\ket{\phi_1}$.  This probability is simply $\braket{h_0|h_0} = q/4 = \eta$.  We may now use Equation \ref{eq:bound-equal-f00f11} to lower bound the quantity $|\braket{f_{0,0}|f_{1,1}}|^2$.

\begin{figure}\center
\includegraphics[width=250pt]{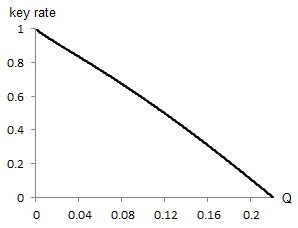}
\caption{A graph of our key rate lower bound when the server $C$ is semi-honest and there is a noisy quantum channel, modeled by two independent depolarization channels, connecting the users.  This graph assumes the noise in the forward direction (when the qubits travel from $C$ to $A$ and $B$) is equal to the noise in the reverse (when qubits return from $A$ and $B$ to $C$).  Observe that the key rate remains positive for all $Q \le 22.05\%$, where $Q$ is the probability that $A$ and $B$'s measurement results differ.}\label{fig:keyrate-semihonest}
\end{figure}

Using these values and Equation \ref{eq:key-rate-bound}, we see that the key rate of this mediated protocol, in the event $p=q=2Q$ (i.e., the noise in the forward and reverse channels is equivalent and the probability that $A$ or $B$'s measurement results are wrong is $Q$), remains positive for all $Q \le 22.05\%$ as shown in Figure \ref{fig:keyrate-semihonest}.  This is an improvement over our original bound of $Q \le 19.9\%$.

Note that, to achieve this high tolerance level in the presence of third-party eavesdroppers, the classical channel connecting $C$ to $A$ or $B$ needs to be authenticated.  This requirement is not necessary in the next scenario we consider where $C$ is fully adversarial.

\subsection{Adversarial Server, Symmetric Attack}
Of course the above section assumed $C$ was semi-honest - that is he followed the protocol exactly.  Now, we assume the server $C$ is adversarial.  He is allowed to prepare any state he likes on step (1) of the protocol, and he may perform any arbitrary operation (allowed by the laws of quantum physics) when the qubits return to him, sending any message of his choice based on his operation.  The only assumption we make in this section is that $C$'s attack is symmetric in that it can be parameterized by only a few parameters (to be discussed).  We make this assumption first, so that we can compare our new key rate bound with our old one from \cite{SQKD-MultiUser} (which assumed a symmetric attack); and second, so that we can better visualize the key rate bound by reducing the many parameters to only a few.  Note that in our original proof of security in \cite{SQKD-MultiUser}, we assumed throughout that $C$'s attack was symmetric (the key rate bound we derived there was based on this assumption).  In our new bound in this paper, however, we made no such assumptions - Equation \ref{eq:key-rate-bound} works even in the most general case.  So, while assuming a symmetric attack is not required for our new proof in this paper, it is easier to visualize our key rate bound and it also allows us to compare our new results with our old.

The first assumption we make is that we may parameterize the values $|\alpha_{i,j}|^2$ using only a single parameter $Q$.  Namely, we have:
\begin{align*}
|\alpha_{0,0}|^2 &= \frac{1-Q}{2} = |\alpha_{1,1}|^2\\
|\alpha_{0,1}|^2 &= \frac{Q}{2} = |\alpha_{1,0}|^2.
\end{align*}

Next, we assume that the probability that $C$ sends ``$-1$'' in the event $A$ and $B$ measure $\ket{0,0}$ is equal to the probability he sends that same message if they both measure $\ket{1,1}$.  Similarly for the case $\ket{0,1}$ and $\ket{1,0}$.  That is:
\begin{align*}
\braket{f_{0,0}|f_{0,0}} &= \braket{f_{1,1}|f_{1,1}} = \mathcal{F}_{=}\\
\braket{f_{0,1}|f_{0,1}} &= \braket{f_{1,0}|f_{1,0}} = \mathcal{F}_{\ne}
\end{align*}
We make no assumptions regarding the comparative relationship between $\mathcal{F}_=$ and $\mathcal{F}_{\ne}$.

Using similar arguments as in \cite{SQKD-MultiUser} (in particular see Equation 25 from that source, along with its derivation), we may find the following upper-bound on $\braket{h_0|h_0}$:
\[
\braket{h_0|h_0} \le \left(\frac{ \sqrt{1-Q}\left(\sqrt{Q\mathcal{F}_{\ne}} + \sqrt{p_w} \right) } {1-Q}\right)^2 = \eta,
\]
where $p_w$ is the probability that $C$ sends the wrong message (namely ``$-1$'') in the event both $A$ and $B$ reflect.

This is our value for $\eta$ allowing us to bound $|\braket{f_{0,0}|f_{1,1}}|^2$ using Equation \ref{eq:bound-general-f00f11} from the previous section.  However, to evaluate this bound, we must determine what to set $\mathcal{F}_=$ and $\mathcal{F}_{\ne}$ to.  Naturally, in practice, these values are simply observed by $A$ and $B$.  However, for this paper, to evaluate our bound, we will consider two examples.

First, in order to compare with our previous bound from our original paper, we will use $\mathcal{F}_{\ne} = Q$ and:
\[
\mathcal{F}_= = \frac{\tilde{p}_a - Q^2}{1-Q},
\]
where $\tilde{p}_a$ is the desired probability of acceptance.  It is trivial to check that, when setting these values thusly, we have $p_a = \tilde{p}_a$ (see Equation \ref{eq:pa}).  We have to set this parameter this way since, in our original proof, we evaluated our key rate equation assuming $p_a$ was $0.5$, then $0.4$, and finally $0.3$ (our original proof did not establish a clear relationship between the value of $p_a$ and the other observed parameters as we did in this new proof).

\begin{figure}\center
\includegraphics[width=250pt]{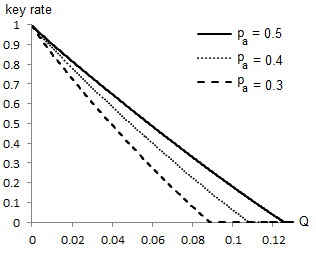}
\caption{Showing a graph of our key rate lower bound when the server is fully adversarial and when $C$'s attack is such that $p_a$, the probability that any particular iteration is accepted by $A$ and $B$, is equal to $0.5$, $0.4$, and $0.3$.  Here, $Q$ represents the probability that $A$ and $B$'s measurement results are different.  This is done to compare with our old bound from \cite{SQKD-MultiUser} (see text).  Indeed, we see that in all three cases, our new bound is superior.  For instance, when $p_a = .5$, our new bound remains positive for all $Q \le 12.5\%$ whereas our old bound from the previous paper remained positive only for $Q \le 10.65\%$.  The difference is even more noticeable for smaller values of $p_a$ as mentioned in the text.}\label{fig:keyrate-adv-1}
\end{figure}

Using these values, we see that, when $Q = p_w$ and $\tilde{p}_a = .5$ the key rate expression is positive for all $Q \le 12.5\%$; when $\tilde{p}_a = .4$ it is positive for $Q \le 10.8\%$; and for $\tilde{p}_a = .3$ it remains positive for $Q \le 8.86\%$.  See Figure \ref{fig:keyrate-adv-1}.  Compare this with our old bound from \cite{SQKD-MultiUser} - there, when $\tilde{p}_a = .5$ our old bound remained positive for $Q \le 10.65\%$, while for $\tilde{p}_a = .3$, the old bound was positive only for $Q \le 5.25\%$.

The second example we consider is based on values determined from the depolarization channel example in the previous section.  We will assume the noise in both channels is equal.  In that case, we have $\mathcal{F}_{\ne} = Q/2$ while $\mathcal{F}_= = 1/2 - Q/2$ (see Equations \ref{eq:dc:f00} and \ref{eq:dc:f01} and recall that $Q = 2p = 2q$).

\begin{figure}\center
\includegraphics[width=250pt]{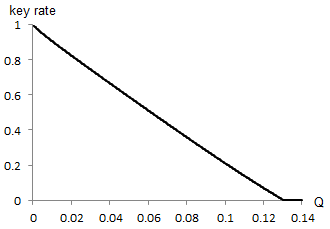}
\caption{Showing a graph of our key rate lower bound when the server is fully adversarial and when $C$'s attack is such that certain parameters agree with the depolarization example (see text).  Here, $Q$ represents the probability that $A$ and $B$'s measurement results are different.  Note that the key rate remains positive for all $Q \le 13.04\%$.}\label{fig:keyrate-adv-2}
\end{figure}

Using these values, we see that, when $Q = p_w$, the key rate remains positive for all $Q \le 13.04\%$.  See Figure \ref{fig:keyrate-adv-2}.

For comparison, BB84 can tolerate up to $11\%$ error while the six-state BB84 can tolerate up to $12.6\%$ (both bounds without pre-processing since we have not considered pre-processing in our proof) \cite{QKD-renner-keyrate,QKD-renner-keyrate2}.  B92 \cite{QKD-B92} was recently shown to tolerate up to $6.5\%$ error assuming a depolarization channel \cite{QKD-B92-Improved}.

We stress that, in practice, there is no need to make such assumptions to determine $\mathcal{F}_=$ and $\mathcal{F}_{\ne}$: these are parameters that are observed directly.  We made assumptions only to visualize our key rate bound and to compare it with our previous work.

\section{Conclusion}
We have provided a new proof of security for the mediated semi-quantum key distribution protocol presented in \cite{SQKD-MultiUser}.  While a proof of security was provided in that original source, we have demonstrated that our new key rate bound provides a more optimistic rate.  Indeed we see that in every scenario considered, our new proof demonstrates that the tolerated noise level of the protocol (the maximal amount of noise before $A$ and $B$ should abort) is strictly larger than the old key rate bound.  Our new proof also does not make any assumptions about the attack used by $C$.  Note that, while we only evaluated our key rate bound in two specific examples - the semi-honest case and the symmetric adversarial case - our proof works in even the most general of scenarios (i.e., a non-symmetric adversarial attack).  Our work in this paper has shown that this mediated SQKD protocol can tolerate noise levels surpassing the maximal tolerated noise thresholds of many other fully quantum protocols.  Finally, the proof technique we used may hold utility in security proofs of other protocols (quantum or semi-quantum) which rely on a two-way quantum communication channel.

Many open problems remain.  Most important, perhaps, is that we considered only the perfect qubit scenario; dealing with problems such as multi-photon attacks is an interesting area of research (in any two-way quantum protocol).  Also, can a mediated SQKD protocol be designed that is more efficient - recall that, in the absence of noise and with an honest server, only half of the sent qubits can contribute to the raw key.  Finally, can a mediated SQKD protocol be designed without the need for $C$ to prepare and measure in the Bell basis.  As mentioned in \cite{SQKD-MultiUser}, the answer to this seems to be positive, however the security proof is more involved.  Perhaps the techniques we developed and used in this paper can be adapted towards this goal.

%%\bibliographystyle{unsrt}
%%\bibliography{QC_Bib,QKD_Bib,SQKD_Bib}

\end{document}